\documentclass[12pt,preprint]{aastex}
\slugcomment{To appear in the Astrophysical Journal}

\lefthead{Sokal et al.}
\righthead{Berkeley 87}

\begin{document}

\title{{\em Chandra} Detects the Rare Oxygen-type Wolf-Rayet Star WR 142 
and OB Stars in  Berkeley 87}

\author{Kimberly R. Sokal\footnote{CASA, Univ. of Colorado, Boulder, CO, USA 80309-0389; 
kimberly.sokal@colorado.edu, stephen.skinner@colorado.edu} ,  
Stephen L. Skinner\altaffilmark{1}, 
Svetozar A. Zhekov\footnote{JILA, Univ. of Colorado, Boulder, CO, USA 
80309-0440} $^,$\footnote{On leave from Space Research Institute, Sofia, Bulgaria} ,
Manuel  G\"{u}del\footnote{Dept. of Astronomy, Univ. of Vienna,
T\"{u}rkenschanzstr. 17,  A-1180 Vienna, Austria} , and 
   Werner Schmutz\footnote{Physikalisch-Meteorologisches Observatorium 
   Davos and World Radiation Center 
(PMOD/WRC), Dorfstrasse 33, CH-7260 Davos Dorf, Switzerland}}

\newcommand{\ltsimeq}{\raisebox{-0.6ex}{$\,\stackrel{\raisebox{-.2ex}
{$\textstyle<$}}{\sim}\,$}}
\newcommand{\gtsimeq}{\raisebox{-0.6ex}{$\,\stackrel{\raisebox{-.2ex}
{$\textstyle>$}}{\sim}\,$}}

\begin{abstract}
We present first results of a {\em Chandra}  X-ray observation
of   the  rare oxygen-type Wolf-Rayet star WR 142 (= Sand 5 = St 3) harbored in 
the young, heavily-obscured cluster Berkeley 87. Oxygen type WO stars are thought to be the 
most evolved of the WRs and progenitors of supernovae or gamma ray bursts. As part of an X-ray 
survey of supposedly single Wolf-Rayet stars, we observed WR 142
and the surrounding Berkeley 87 region with {\em Chandra} ACIS-I. We detect WR 142 as a
faint, yet extremely hard X-ray source. Due to weak emission, its nature as a thermal or 
nonthermal emitter 
is unclear and thus we discuss several emission mechanisms. Additionally, we report 
seven detections and eight non-detections by {\em Chandra} of massive OB stars in Berkeley 87, two of which are
bright yet soft X-ray sources whose spectra provide a dramatic contrast to the hard 
emission from WR 142.

\end{abstract}


\section{Introduction}

Wolf-Rayet (WR) stars  are massive, highly-evolved stars nearing the end of their lives as 
supernovae (SN) or as collapsing objects emitting gamma-ray bursts (GRB, e.g. 
MacFadyen \& Woosley 1999, Postnov \& Cherepashchuk 2001, Georgy et al. 2009). WR 
stars undergo  rapid mass loss through strong winds, with initial masses of 
$\geq$25M$\sun$ (Crowther 2007). The classification of WR stars is determined spectroscopically in the 
optical and is divided among the nitrogen-rich WNs, carbon-rich WCs, and oxygen-rich 
WOs. For the most part, single WR stars are thought to follow the evolutionary path: O 
$\rightarrow$ (LBV/RSG) $\rightarrow$ WN $\rightarrow$ WC $\rightarrow$ WO 
$\rightarrow$ SN [or GRB] (Conti et al. 1983, Crowther 2007). The initial mass of the O star 
determines whether it passes through an intermediate luminous blue variable (LBV) or red 
supergiant (RSG) phase (Crowther 2007). 

Sanduleak (1971) noticed a class of stars that did not have planetary nebulae (PN) but 
displayed a WR-like spectrum with strong O VI doublet emission similar to those found in 
the central stars of PN named the O VI sequence by Smith and Aller (1969). These stars 
were suggested to be a separate WO sequence of Wolf-Rayet stars (rather than an 
extension of the WC sequence) by Barlow \& Hummer (1982). WO stars are thought to be in 
the late helium-burning stage, or possibly the carbon-burning stage, where the enhanced 
oxygen abundance compared to less evolved WR stars is revealed by mass loss stripping 
(Barlow \& Hummer 1982). Of the 298 galactic WR stars in the appended VIIth catalogue of 
galactic Wolf-Rayet stars (van der Hucht 2001, 2006), only four are of the rare WO spectral 
type, including the star WR 142, also known as Sand 5 (Sanduleak 1971) and St 3 
(Stephenson 1966). Crowther et al. (1998) developed a new classification scheme using 
primary and secondary oxygen line ratios that confirmed previous classifications of WR 142 
as a member of the subclass WO2 (Barlow \& Hummer 1982, Kingsburgh et al. 1995). 

Despite new discoveries of WR stars in the galactic plane, including a WO type as the 
exciting star of the planetary nebula Th 2-A (Weidmann et al. 2008), little is yet understood 
about the mechanisms that drive high-energy processes such as X-ray emission in single 
WR stars, particularly so for the rare WO stars. Although WR stars have much stronger 
winds and are more evolved chemically, the line-driven instabilities that are thought to give 
rise to soft X-rays (kT $<$ 1 keV)  from  shocks in O star winds may also be present in WR 
winds (Gayley \& Owocki 1995). If that is the case, then WR stars may also be capable of 
producing soft X-rays from radiative wind shocks (Baum et al. 1992).

Few single WR stars had been studied with high sensitivity in X-rays until recently. 
Thus far, several apparently single WN stars have been observed to emit X-rays (Skinner et 
al. 2002, Ignace et al. 2003, Oskinova 2005). 
An ongoing  {\em Chandra} and {\em XMM-Newton}  survey has recently 
detected the apparently single WN stars  
WR 2, WR 18, WR 79a, and WR 134 with X-ray 
luminosities log $L_{\rm X}$ $\approx$ 32.2 - 32.7 erg s$^{-1}$
(Skinner et al. 2010b), 
comparable to some WR$+$OB binaries like $\gamma^2$ Vel (WC8$+$O7) with log 
L$_{\rm X}$ $=$ 32.9 erg s$^{-1}$ (Skinner et al. 2001) or WR 147 (WN8$+$OB) with log 
L$_{\rm X}$ $=$ 32.83 erg s$^{-1}$ (Skinner et al. 2007; see also Sec. 4.4). Only upper limits 
of log L$_{\rm X}$ $<$ 29.82 - 30.97 erg s$^{-1}$ exist from observations of single WC stars 
(Oskinova et al. 2003, Skinner et al. 2006). The WC sequence is thus either faint in the X-rays
 or possibly X-ray quiet. WR 142 just recently became the first WO star to be detected in 
the X-rays using {\em XMM-Newton} (Oskinova et al. 2009). 

WR 142 resides in the open cluster Berkeley 87. Table 1
summarizes the general properties of WR 142. Berkeley 87  lies
in a heavily obscured region of the 
Milky Way in Cygnus. Initial cluster age estimates of 
$\sim$ 1 - 2 Myr (Turner \& Forbes 1982, hereafter TF82) have now been revised upward. A 
study restricted 
to three of the highest mass cluster stars suggests a slightly older 
age of $\sim$ 3 Myr (Massey et al. 2001). A more recent study (Turner et al. 2010) gives 
$\sim$ 5 Myr. Berkeley 87 is 
an interesting cluster containing $\approx$ 105 cluster members 
identified by an optical study (TF82),  B-supergiants including HD 229059, an O8.5-O9 star 
BD+36 4032,
a possible luminous blue variable Be star V439 Cyg, 
and the pulsating M3-supergiant BC Cyg. Additionally, 
OH masers and compact HII regions trace massive star 
formation 3$'$ - 9$'$ north of WR 142 
(Argon et al. 2000, Matthews et al. 1973). At a distance 
of 1230 $\pm$ 40 pc (Turner et al. 2006), the
proximity of Berkeley 87 offers ample opportunity 
to learn about the properties of rare objects 
such as WR 142 as well as the surrounding cluster.

As part of an X-ray survey aimed at determining if single Wolf-Rayet stars that are not 
known to be in binary systems are X-ray emitters (specifically which spectral types), we 
observed  WR 142 with {\em Chandra}. We are unaware of any evidence pointing to a 
companion of WR 142. WR 142 was chosen over other WO stars because of its relatively 
nearby distance,  low A$_{\rm V}$ compared to other WR stars, and interesting surrounding 
cluster. Its astonishing supersonic wind (v$_{\infty}$ $=$ 5500 km s$^{-1}$, Kingsburgh et 
al. 1995) and an unusual optical detection of diffuse emission from the C IV doublet 
(Polcaro et al. 1991) that is a signature of shocked gas make WR 142 an even more 
interesting target as one of the few oxygen-type WR stars.

The primary objective of the study presented here is to determine the X-ray properties of 
WR 142. X-ray emission from WR 142 was previously observed by {\em XMM-Newton} 
(Oskinova et al. 2009), but {\em Chandra} has several advantages over 
{\em XMM-Newton}, including improved angular resolution (by a factor of $\sim$ 4 - 5), sharp on-axis 
point-spread function (PSF) to provide checks on source extent, longer uninterrupted 
exposure that provides a continuous light curve and more stringent test for variability, and 
lower instrumental background. Such benefits allow for reliable source identification, 
especially when compared with optical positions; negligible background subtraction for 
lightcurves and spectra; and searches for the presence of diffuse shock emission around 
WR 142. We analyze the faint X-ray spectrum of WR 142 and discuss 
possible X-ray emission mechanisms. We also report {\em Chandra} detections of seven 
other massive OB stars in Berkeley 87 and compare the much softer spectra from two X-ray 
bright OB stars to the hard spectrum of WR 142. In this paper, we focus on WR 142 and 
massive OB stars in Berkeley 87. A first overview of the {\em Chandra} results for the cluster 
as a whole was presented by Skinner et al. (2010a) and the cluster will be discussed in more detail in a future paper.

\section{Chandra Observations and Data Reduction}

The {\em Chandra} observation (ObsID 9914) centered near WR 142 began on
2009 February 2 at 02:41:24 TT and ended on February 2 at 22:52:55 TT. The nominal 
pointing coordinates were located 28$''$ NE of WR 142 at R.A. $=$ 20{\em h} 21{\em m}  
46.42{\em s}, Decl. $=$ $+$37$^{\circ}$ 22$'$ 44.9$''$  (J2000.0). The observation was obtained with 
the ACIS-I (Advanced CCD Imaging Spectrometer) imaging array in faint 
timed-event mode with 3.2 s frame times. The exposure live time was  70148 s. ACIS-I 
consists of four front-illuminated 1024 $\times$ 1024 pixel CCDs
with pixel size $\approx$ 0.492$''$ and has a combined field of view (FOV) of $\approx$ 
16.9$'$ $\times$ 16.9$'$.
More information on 
{\em Chandra} and its instrumentation can be found in the {\em Chandra} Proposer's 
Observatory Guide (POG)\footnote {See http://asc.harvard.edu/proposer/POG}.

The {\em Chandra} X-ray
Center (CXC) provided a Level 2 events file, which was  analyzed with CIAO version 
4.1.2\footnote{Further information on 
{\em Chandra} Interactive
Analysis of Observations (CIAO) software can be found at
http://asc.harvard.edu/ciao.} using standard science 
threads. To determine the source centroid position, the net counts, and a 3$\sigma$ 
extraction region of the target WR 142, 
we used the CIAO {\em wavdetect} tool.  We ran {\em wavdetect} on full-resolution 
images using events in the 0.3 - 8 keV
range, chosen to reduce background. The {\em wavdetect}
threshold was set at $sigthresh$ $=$ 1 $\times$ 10$^{-6}$. 
Scale sizes of 1, 2, 4, 8, and 16 were used.  The 3$\sigma$ extraction region determined 
for WR 142 by {\em wavdetect} was an ellipse with an area of $\approx$ 11.25 pixels$^{2}$ (Figure 1) 
and was used for spectral and timing analysis.  The CIAO task {\em srcextent}, which 
utilized point-spread-function (PSF) information, was also used to determine the source 
size of WR 142 and determine if its emission is point-like or extended.

Within the 0.3 - 8 keV energy range, light curves were extracted using CIAO {\em 
dmextract} and source variability was tested by two different methods. Using unbinned 
photon arrival times, we applied the Kolmogorov-Smirnov (KS) test (Press et al. 1992). For 
comparison, the Gregory-Loredo algorithm (Gregory \& Loredo 1992, 1996) was 
applied by using CIAO {\em glvary}\footnote{http://cxc.harvard.edu/ciao/ahelp/glvary.html.} 
tool. 

CIAO {\em specextract} was used to extract  source and background spectra. The tool {\em 
specextract} also
created response matrix files (RMFs) and auxiliary
response files (ARFs). Spectral analysis utilized the HEASOFT {\em Xanadu}\footnote{The 
XANADU X-ray analysis software package is developed and maintained by NASA's High 
Energy Astrophysics Science Archive Research Center (HEASARC). See 
http://heasarc.gsfc.nasa.gov/docs/xanadu/xanadu.html for further information.} software package 
including XSPEC vers. 12.5.0. The background region was chosen to be a source-free 
annulus centered on WR 142 of r $=$ 4 to r $=$ 16 pixels. From this region, we estimate 
$<$ 1 background count (0.3 - 8 keV) inside the 3$\sigma$ elliptical source region for WR 
142. The background inside the source region is negligible relative to the number of 
source counts. 

\section{Results}      
   
\subsection{Imaging and Source Identification}

WR 142 is detected by {\em Chandra} as an X-ray source at J202144.35$+$372230.7 
(Figure 1). Table 2 summarizes some properties of the X-ray emission from WR 142. 
These coordinates are the source centroid as determined by {\em wavdetect} using an 
energy filtered image (0.3 - 8 keV). Without further astrometric calibration, the {\em 
Chandra} ACIS-I absolute position has an accuracy of $\approx$ 0.4$''$ 
(68\% limit)\footnote{http://cxc.harvard.edu/cal/ASPECT/celmon/.}. The X-ray position of WR 142 is in 
exact agreement with the 2MASS position of J202144.35$+$372230.7. Additionally, the 
{\em Chandra} position has offsets from 
optical counterparts of only 0.20$''$ from the USNO-B1.0 source at 
J202144.35$+$372230.9 and 0.23$''$ from the 
HST GSC2.3.2 source at J202144.36$+$372230.5. All offsets are well within the positional 
accuracy of {\em Chandra}. 

Results from the CIAO tool {\em srcextent} indicate WR 142 is not extended. Use of a PSF 
file specifically tailored for our observation with the WR 142 spectrum and the position of 
WR 142 on the ACIS-I detector yields a source size of only 0.52$''$ (0.41-0.64$''$, 90$\%$ 
confidence), equivalent to about one ACIS pixel, and a PSF size of 0.85$''$ (0.66-1.03$''$, 
90$\%$ confidence). Thus, the X-ray emission of WR 142 is not extended at the 90\% 
confidence level. 

\subsection{Spectral Analysis}

The X-rays from WR 142 are weak but extremely hard, with a mean photon energy of 4.31 
keV (Table 2). Surprisingly, WR 142 does not exhibit any significant emission below 2 keV 
and there is no evidence of a soft component (Figures 2, 3). We find only one event below 2 
keV, with an energy of 1.89 keV. This is in contrast to a faint soft component
    at E $\leq$ 1 keV  in the {\em XMM-Newton}
    EPIC MOS spectrum of WR 142 reported by
    Oskinova et al. (2009). Assuming that
    no variability occurred in the WR 142 source
    spectrum or N$_{\rm H}$ between the two
    observations, the most likely explanation
    for this difference would be the much
    lower effective area of  {\em Chandra}
    ACIS-I at low energies E $\leq$ 1 keV.

We tried to fit the spectrum of WR 142 (Figure 3) with several different models: a 1T (one 
temperature) APEC optically thin plasma model and a power-law model. The APEC model 
requires a very high temperature and thus a thermal bremsstrahlung model is similar. 
Because of low counts, the spectral fit results were somewhat sensitive to binning strategy. 
Representative spectral fits are shown in Table 3, based on fits of unbinned spectra. 
Between the different models, there is little variation in the quality of fit using 
C-statistics (Cash 1979). However, determining whether a particular
model is statistically acceptable using C-statistics is not as straightforward as with $\chi^2$ 
statistics (Heinrich 2003).

Allowing the abundances to vary from solar displays no significant improvement to the APEC 
fits, nor does the use of representative WO abundances (Table 3). This is largely because 
we do not  detect any emission lines in the spectrum from elements such as O which are 
highly nonsolar in WO stars. Additionally, all of the O and C lines that lie in the {\em 
Chandra} bandpass are below 1 keV where we do not detect any signal. 

The data show a slight preference for a model that includes a gaussian line for an iron line in 
the $\approx$ 6.4 - 6.67 keV energy range (Table 3). The counts from WR 142 are not 
sufficient to distinguish between the possible iron lines. The event energies closest to the 
possible iron emission, 6.24, 6.40, 6.60, and 6.63 keV, show two events within 0.1 keV of the 
iron K line complex at 6.67 keV (Fe XXV He-like triplet). If a weak Fe line is present, it is likely 
to be the 6.67 keV iron K line, which has been detected in other WR stars.

We have also experimented with fits of binned spectra using $\chi^2$ statistics in order to 
estimate confidence bounds on important fit parameters (F$_{\rm X}$, N$_{\rm H}$, kT, 
$\Gamma_{\rm pow}$). Given the limited number of counts (46), binning at desired  levels of 
$>$10 counts per bin for fitting with $\chi^2$ statistics was not feasible. We thus consider 
spectra binned to a minimum of 6 counts per bin. Results indicate that if the plasma is 
thermal, as the APEC models assume, then the temperature kT must be very high, with a 
lower 90\% confidence bound of kT $>$ 3.6 keV (Table 3). The {\em Chandra} data do not 
place a useful upper bound on kT.

In determining the X-ray flux, the unbinned spectral fits display fairly consistent results. The fit 
with the lowest C-statistic shown in Table 3 gives an observed (absorbed) X-ray flux of 
F$_{\rm X}$ $=$ 1.8 (0.8 - 2.5) $\times$ 10$^{-14}$ erg s$^{-1}$ cm$^{-2}$ (0.3 - 8 keV, 
1$\sigma$ confidence intervals). The error bounds reflect the range of fluxes measured for the 
different models in Table 3. The best fit models give X-ray luminosities of log L$_{\rm X}$ 
$=$ 30.72 - 30.82 (ergs s$^{-1}$), which are derived from unabsorbed fluxes (0.3 - 8 keV) 
resulting from different unbinned spectral fits in Table 3. 

All indications are that the absorption towards WR 142 is high, with N$_{\rm H}$ around 
4 - 5 $\times$ $10^{22}$ cm$^{-2}$. This is about 3 - 4 times higher than the absorption estimated 
from the observed optical extinction of 1.3 $\times$ 10$^{22}$ cm$^{-2}$ obtained using 
A$_{\rm V}$ $\approx$ 6.1 (van der Hucht 2001, Gorenstein 1975 conversion). Spectral fits of 
WR 142 with N$_{\rm H}$ fixed at the value determined from optical extinction give much 
larger fit residuals than the fits with larger 
absorption given in Table 3.

\subsection{Timing and Variability Analysis}

The source shows no obvious large-scale variations nor flares in its X-ray lightcurve (Figure 
4). As for short term variability, KS statistics and the Gregory-Loredo algorithm give a 
probability of constant count rate of 0.77 and 0.40, respectively. Thus, neither test implies 
variability at high confidence levels ($>$90\%) on timescales of hours during the $<$1 day 
{\em Chandra} observation. 

For purposes of comparing the observed X-ray flux by {\em Chandra} and the previously 
published {\em XMM-Newton} X-ray flux of F$_{\rm X}$ $=$ 4 $\pm$ 2 $\times$ 10$^{-14}$ 
erg cm$^{-2}$ s$^{-1}$ from 0.25 - 12 keV (Oskinova et al. 2009), we measure F$_{\rm X}$ 
$=$ 3 $\pm$ 1 $\times$ 10$^{-14}$ erg cm$^{-2}$ s$^{-1}$ (0.25 - 12 keV) using an average 
value of the fluxes from the extended energy range from the three models shown in Table 3. 
Thus, {\em Chandra} and {\em XMM-Newton} X-ray fluxes agree to within the uncertainties. 
Based on the above flux comparison, any variability over the approximate one year interval 
between the {\em Chandra} and  {\em XMM-Newton} observations was a factor of $\sim$ 2 or 
less.

\subsection{Summary of WR 142 X-ray Properties}
Based on the {\em Chandra} data discussed above, the X-ray
properties of WR 142 can be summarized as follows: 
(i) source structure is consistent with a point source at
{\em Chandra}'s angular resolution, (ii)  no significant variability
is detected over a timescale of $\approx$ 1 day, 
(iii)  strong X-ray absorption below 2 keV, with an absorption 
column density (N$_{\rm H}$) that is a factor of $\approx$ 3 - 4 greater than 
expected from published optical estimates, (iv) no evidence for a 
soft X-ray component below 2 keV, (v)  a spectrum dominated by
hard X-rays above 2 keV that can be fit equally well using a 
very high temperature thermal plasma model or a power-law
model, and (vi) a relatively low X-ray luminosity 
log L$_{\rm X}$(0.3 - 8 keV) $=$ 30.7 - 30.8 ergs s$^{-1}$
(at d $=$ 1.23 kpc), which is $\approx$ 10 times less than
found for two X-ray bright OB stars in Berkeley 87 (Section 4). We discuss these 
properties in more detail below.

\section{OB Stars in Berkeley 87}

In this section, we comment briefly on OB stars in Berkeley 87 contained in 
the {\em Chandra} FOV. We report {\em Chandra} detections of seven OB 
stars and identify eight undetected B-type stars (Table 2) that are likely 
cluster members (TF82). Figure 1 shows the relative positions of the X-ray 
bright stars and the noteworthy non-detection of the LBV candidate V439 
Cyg.

\subsection{BD$+$36 4032 }

This star was classified as O9V by TF82 (Berkeley 87-25)
and was the only object explicitly identified as an O-type
star in their catalog of 105 likely members of
Berkeley 87. The spectral type was later refined to O8.5 III
by Massey et al. (2001), who also estimated the mass to
be M$_{*}$ $=$ 39 M$_{\sun}$. BD$+$36 4032 lies $\approx$ 3$'$
northwest of WR 142. 

X-ray characteristics are listed in Table 2. 
The X-ray spectrum is soft (Figure 3) with almost all of
the emission emerging below 2 keV, which is dramatically different than 
the hard emission from WR 142. Reasonably good fits
can be obtained using a solar abundance 1T APEC model with a cool 
plasma
component at kT $\approx$ 0.6 keV (Table 4).  The 1T APEC fit can be 
improved slightly by adding
a second temperature component between 1 - 2 keV, 
but almost all of the emission measure still resides
in the cool component.

The best-fit N$_{\rm H}$ given in Table 4 equates to 
A$_{\rm V}$ $=$ 5.1 [4.4 - 6.5] mag (Gorenstein 1975).
This is in good agreement with the optically-determined
value A$_{\rm V}$ $=$ 4.8 mag using E(B $-$ V) $=$ 1.60
and A$_{\rm V}$ $=$ 3E(B $-$ V) (TF82). Thus, the 
excess X-ray absorption detected in WR 142 is
not present toward this O star.

\subsection{HD 229059}

This B supergiant has V $=$ 8.71 mag and was the brightest  member of
Berkeley 87 listed by TF82 (Berkeley 87-3). They assigned a spectral type
B2 Iabe but Massey et al. (2001) classified it as  B1 Ia.

The {\em Chandra} spectrum of  HD 229059 is nearly identical
to that of  BD$+$36 4032 (Figure 3). One noticeable difference
is that  HD 229059 shows stronger emission in the 1.8 - 1.9 keV
range, which is likely due to stronger line emission  
from  the Si XIII triplet (1.84 - 1.87 keV).

Spectral fits of  HD 229059 give nearly identical results  
to BD$+$36 4032. Table 4 summarizes the best-fit
1T APEC model.
No significant fit improvement was obtained by adding
a second temperature component.
The N$_{\rm H}$ from Table 4 gives A$_{\rm V}$ $=$ 5.2 [4.5 - 5.6] mag,
in excellent  agreement with the optically-determined  value 
A$_{\rm V}$ $=$ 5.1 mag (TF82). Thus, we find no excess X-ray
absorption toward HD 229059.

\subsection{Berkeley 87-4}

This B0.2 III star lies $\approx$ 5.1$'$ west of WR 142 and is object
number 4 in the list of likely cluster members compiled by TF82. It
was included in the study of Massey et al. (2001)  who obtained a
mass estimate M$_{*}$ $=$ 23 M$_{\sun}$. Berkeley 87-4 was the third brightest
{\em Chandra} X-ray detection of the OB stars in Berkeley 87 (Table 2).
There are sufficient counts for a crude spectral fit, and a 
1T APEC model  gives N$_{\rm H}$ and kT values that are 
similar to the brighter OB stars BD$+$36 4032 and HD 229059
but a lower X-ray luminosity (Table 4). The X-ray derived N$_{\rm H}$
agrees to within the model uncertainties with that expected from 
optical reddening (TF82).

\subsection{V439 Cyg}

This variable star is  also known as
MWC 1015 and is object number 15 in the list of
likely cluster members compiled by TF82.
V439 Cyg has anomalous colors and is displaced
well to the redward side of the zero-age
main-sequence in the cluster color-magnitude
diagram (TF82). The star's spectrum is peculiar 
and evidently variable, with most recent studies 
assigning an early B spectral type of B0e
(Polcaro \& Norci 1998) or B[e] (Massey et al. 2001).
V439 Cyg shows some similarities to a
LBV (Polcaro \& Norci 1998).

V439 Cyg lies $\approx$ 3.2$'$ NW of WR 142 and is
a bright near-IR source with 2MASS data 
giving K$_{\rm s}$ $=$ 7.59 mag and position
2MASS J202133.57$+$372451.57.  There is no X-ray
source at or  near this position in our 
{\em Chandra} image. A circular region of radius
1.96$''$ (90\% encircled energy at E $=$ 2 
keV\footnote{http://cxc.harvard.edu/cal/Acis/Cal\_prods/psf/eer\_on.html.})
centered on the V439 Cyg 2MASS position
encloses only one photon  in the ACIS-I image
(0.3 - 8 keV range).

To compute an X-ray upper limit, we assume a
conservative 4-count ACIS-I detection threshold 
and an intrinsic thermal stellar spectrum similar
to the two OB stars discussed above. Using
N$_{\rm H}$ $=$ 1.0 $\times$ 10$^{22}$ cm$^{-2}$
corresponding to E(B$-$V) $=$ 1.53 (TF82) 
and a Raymond-Smith solar-abundance thermal 
plasma model with kT $=$ 0.6 keV, the {\em Chandra} 
PIMMS\footnote{For information on PIMMS (Portable
Interactive Multi-Mission Simulator) see
http://cxc.harvard.edu/ciao/ahelp/pimms.html.}
simulator gives an unabsorbed flux upper limit
F$_{\rm X}$(0.3 - 8 keV) $\leq$ 
3.9 $\times$ 10$^{-15}$ ergs cm$^{-2}$ s$^{-1}$.
If V439 Cyg is a cluster member, then d $=$ 1.23 kpc
gives log L$_{\rm X}$ $\leq$ 29.85 (ergs s$^{-1}$).

Given the unusual nature of V439 Cyg, further consideration
of its X-ray non-detection is warranted. In this regard,
it is important to note that {\em Chandra} detected 
only the most luminous  B stars in the cluster having 
absolute magnitudes  m$_{\rm V}$ $\leq$ $-$2.1 (Sec. 4.5). 
Several less-luminous early B stars were undetected (Table 2 notes).
This is likely a consequence of the known L$_{\rm X}$ $\propto$ L$_{\rm bol}$
trend for O and early B-type stars (Bergh\"{o}fer et al. 1997).
The absolute magnitude of V439 Cyg is not well-known due
to uncertain reddening, but it appears to lie at or near
the m$_{\rm V}$ cutoff for our {\em Chandra} detections
(Figure 7 of TF82). Thus, V439 Cyg may have insufficient
L$_{\rm bol}$ to produce an X-ray luminosity above the 
{\em Chandra} detection threshold.

Excess absorption toward V439 Cyg may have also partially 
contributed to its non-detection. The anomalous colors of 
V439 Cyg imply excess reddening, and it may be surrounded 
by a shell of material that was ejected during a previous
mass loss phase (Polcaro \& Norci 1998). A gas-rich shell  
would increase the  line-of-sight photoelectric absorption 
to lower  energy X-rays (E $<$ 1 keV). 

If V439 Cyg is indeed a LBV (Polcaro \& Norci 1998),
it could be evolving into the WN phase. For stars 
of initial masses $\sim$ 40 - 75 M$_{\sun}$, the 
evolutionary sequence is (Crowther 2007):
O $\rightarrow$ LBV $\rightarrow$ WN$_{\rm H-poor}$.
Some putatively single WN7-9 stars (also known as 
WNL stars) have so far gone undetected in X-rays (Figure 5). But a few recent
WNL detections have been reported, including an unambiguous detection of 
the apparently single WN9ha star WR79a
(Skinner et al. 2010b), a possible faint detection of WR 16 (WN8h) in a recent 
{\em XMM-Newton} observation (Figure 5), and a new
{\em Chandra} detection of the WN8 component of the close
WR$+$OB binary system WR 147 (Zhekov \& Park 2010).
Although the nature and evolutionary status of V439 Cyg 
remain somewhat of a mystery, its X-ray non-detection 
would not be in discord with an evolved star that is
approaching the WNL phase, given that a few other WNL stars have also 
gone undetected in X-rays.

\subsection{Other B Stars in Berkeley 87}

In addition to the OB stars discussed above, {\em Chandra}
provides faint X-ray detections of four other B stars in 
Berkeley 87. These are Berkeley 87 no. 13 (B0.5 III), 
no. 24 (B1 Ib), no. 26 (B0.5 Iab), and no. 32 (B0.5 III).
Their basic properties are summarized in Table 2, but
they lack sufficient counts for spectral analysis.
We also report non-detections of seven other B stars
in Berkeley 87 (Table 2 notes). 

The {\em Chandra}
results  show a noticeable trend in that all of
the B star detections have high luminosities 
(m$_{\rm V}$ $\leq$ $-$2.1; TF82 photometry). 
Most, but  not all, of the undetected B stars are 
less luminous (m$_{\rm V}$ $\geq$ $-$2.1). However,
there are a few interesting exceptions. The B1 V
star Berkeley 87-31 (m$_{\rm V}$ $=$ $-$2.7; TF82)
lies nearly on-axis at an offset of 18.7$''$ from
WR 142 but was not detected. However,
it does lie near an ACIS-I CCD gap. Two other luminous 
but undetected B stars, no. 9 (B0.5 V) and no. 38
(B2 III), lie further off-axis (3.9$'$ and 5.0$'$,
respectively).

\section{Discussion}

\subsection{Excess X-ray Absorption Toward WR 142}

{\em Chandra} spectral fits of WR 142 (Table 3) give an absorption column 
density of N$_{\rm H}$
$\approx$ 4 - 5 $\times$ 10$^{22}$ cm$^{-2}$. In contrast, optical studies yield 
N$_{\rm H}$ $\approx$ 1 $\times$ 10$^{22}$ cm$^{-2}$ (TF82, Barlow \& 
Hummer 1982, van der Hucht 2001, Gorenstein 1975 conversions) for 
Berkeley 87 cluster members. Our X-ray analysis of the cluster members 
BD$+$36 4032 and HD 229059 (Sec. 4) also yields 
N$_{\rm H}$ $\approx$ 1 $\times$ 10$^{22}$ cm$^{-2}$.
Thus, the X-ray spectra clearly show that excess absorption
is present toward WR 142 that is not seen toward two other
massive cluster members. This could arise either in the
dense WO wind or perhaps in cold circumstellar gas.

Excess X-ray absorption seen as a disagreement between absorption
based on optical extinction and
the X-ray fit N$_{\rm H}$ has been seen in other 
WR stars as well, such as $\gamma$ Vel$^2$ (Skinner et al. 2001) and 
recently detected WN stars (Skinner et al. 2010b). 

Inhomogeneous extinction, although present, does not account for
the discrepancy in X-ray and optical N$_{\rm H}$ values for WR 142. The 
spread in extinction
values for cluster members is small, from 
N$_{\rm H}$ $=$ (0.8 - 1.3) $\times$ 10$^{22}$ cm$^{-2}$ (TF82, Gorenstein 
1975 conversion). Additionally, the smallest foreground extinction is near the 
south and WR 142 is $\approx$ 2.6$'$ southeast of the cluster center (TF82), 
indicating the excess N$_{\rm H}$ is due to local absorption associated with 
WR 142.

If the excess absorption is due to the wind, we can estimate
the radius at which the X-rays emerge using a wind optical
depth calculation. From the observed {\em Chandra} WR 142 spectrum, the
X-rays are entirely absorbed below $\approx$ 2 keV. Using
generic WO abundances (van der Hucht et al. 1986) and assuming the mass-
loss parameters in
Table 1, the radius of optical depth unity at E $=$ 2 keV is
R($\tau$ $=$ 1, 2 keV) $\approx$ 1500 R$_{\sun}$. Clumping in the WR winds 
may reduce the mass 
loss rate by a factor of 2 - 4 (Crowther 2007) and would reduce R($\tau$ $=$ 1, 2 
keV) by the same factor. 
R($\tau$ $=$ 1) is well outside the wind acceleration zone
and the wind will already have reached terminal speed, assuming
a standard $\beta$ $=$ 0.5 - 1.0 wind velocity law.
It should be kept in mind that this is a {\em minimum} radius
for 2 keV X-rays to escape. 

Another possibility regarding excess absorption is the presence of cold 
circumstellar gas. Nebulosity around WR 142 has been detected (Miller \& Chu 
1993), though a ring nebula is not present. Polcaro et al. (1997) note excess 
reddening surrounding WR 142 of at least 
N$_{\rm H}$ $=$ 2.7 $\times$ 10$^{21}$ cm$^{-2}$ (Gorenstein 1975 
conversion) suggesting dense material, possibly a nebula from the stellar wind. 
Alternatively, Lozinskaya (1991) detected a distant IR shell ($\sim$ 23 pc from 
WR 142) whose dynamical age suggests that it formed before, and therefore 
independently, of the WO wind. Though it is apparent that some circumstellar 
material is present, our X-ray spectral fits have not distinguished between wind 
absorption and absorption by  circumstellar gas far from the star. Both the wind 
and local circumstellar material could contribute to the excess X-ray absorption.

\subsection{Is WR 142 a Binary?}

There is no evidence from ACIS images (or other observations) that WR 142 
is a binary.  For an on-axis point source, the 
{\em Chandra} HRMA and ACIS PSF  encircles 70\% of
the source energy for an angular source radius of
$\approx$ 1.2$''$ at 2 keV and  $\approx$ 1.3$''$
at 6 keV\footnote{http://cxc.harvard.edu/cal/Acis/Cal\_prods/psf/eer\_on.html.}. 
If two X-ray sources are present, their separation must be $<$1$''$.

The WR 142 X-ray luminosity  from this {\em Chandra} observation is much less 
than that of known WR$+$O binaries, which are generally very luminous in the 
X-rays. Due to plasma formed in colliding wind shocks, these systems can have 
luminosities approaching L$_X$ $\approx$ 10$^{33}$ ergs s$^{-1}$, such as 
$\gamma^{2}$ Vel (WC8$+$O7.5III; Skinner et al. 2001, Schild et al. 2004). WR 
142 is at least two orders of magnitude less luminous. 

However, the possibility of an unseen companion remains. 
But, a compact companion seems to be unlikely.
Theoretical models of the WR winds accreting onto neutron stars 
predict high luminosity log L$_{\rm X}$ $>$ 34 (ergs s$^{-1}$) (HD 50896 $=$ EZ 
CMa, Stevens \& Willis 1988). Additionally, a neutron star
would be required to spin fast enough to throw off most of the accreting material 
to remain undetected (propeller effect, e.g. Lipunov 1982). If an optically faint 
lower mass normal stellar companion (rather than a compact object) lies close to 
WR 142, it could easily escape detection.

Currently, the binary frequency for galactic WR stars is 
$\sim$40$\%$ (van der Hucht 2001). Additional searches for binarity are needed for WR 142.
     These could include high resolution infrared imaging
     or radio interferometry, though they would only be capable
     of detecting a companion down to separations of a few tenths
     of an arcsecond. More promising approaches would be to
     search for periodic photometric or spectroscopic variability
     that could signal a closer spectroscopic companion. Spectroscopically, an X-ray emitting companion orbiting
  in the WR wind could reveal itself through phase variability
   of the WR star ultraviolet lines (Hatchett \& McCray 1977).

\subsection{Thermal versus Nonthermal Emission}

Low counts in the WR 142 ACIS-I spectrum result 
in nearly identical fits by either thermal APEC models or 
power-law models. As such, the WR 142 
spectrum cannot sufficiently distinguish between
a very high temperature thermal plasma 
and nonthermal processes e.g. inverse Coulomb scattering due to relativistic 
electrons accelerated in wind shocks  (Chen \& White 1991a). 
In general, other WR stars with higher signal-to-noise spectra 
show detections of emission lines indicative of thermal emission. 
While the goodness-of-fit for spectral models 
slightly improves with the addition of the Fe K line for WR 142, 
there are not enough line counts to qualify 
as a definite line detection. There is no obvious evidence of other thermal 
emission lines, 
such as the S line at 2.46 keV. 
Theoretical models do exist that attempt to account 
for the hard X-ray emission from some early type 
stars by nonthermal processes (Chen \& White 1991a). 
However, if WR 142 were 
a nonthermal X-ray source, it would be the first 
putatively single WR star in that class. The possibility of 
nonthermal emission has been previously mentioned 
for other WR stars such as WR 110 but the evidence 
for nonthermal emission was not conclusive  (Skinner et al. 2002). 

\subsection{Thermal Emission Mechanisms}

Thermal X-ray emission is possible through several processes. Often radiative 
wind shocks are considered, yet with no emission below 2 keV detected from WR 
142, a line-driven flow instability model predicting soft X-rays from radiative wind 
shocks that is typically assumed for O stars (Gayley \& Owocki 1995) cannot 
explain the hard emission. We discuss wind kinetic energy conversion, colliding 
wind shocks, magnetically confined wind shocks (MCWS), and coronal emission 
as possible thermal emission processes. 

\subsubsection{Kinetic Energy of the Wind}

Due to the extreme mass loss rate, the kinetic energy in the wind of WR 142 is 
enough to account for the X-ray luminosity. The kinetic energy of the wind 
supplies log L$_{\rm wind}$ $\approx$ 38.2 (ergs s$^{-1}$) which is much 
greater than the X-ray luminosity observed of log L$_{\rm X}$ $\approx$ 31 (ergs 
s$^{-1}$). The effects of clumping in the wind on the mass loss rate would not 
change this conclusion, reducing log L$_{\rm wind}$ by 0.6 dex (ergs s$^{-1}$) 
at most. Thus even if the efficiency for converting the wind kinetic energy into X-
rays were small, the wind still could supply the energy. 

\subsubsection{Colliding Wind Shocks}

Because WR 142 has no known companion, a colliding wind shock model does 
not obviously apply. However, such models could explain 
the  high temperatures of the X-ray emission produced from the high wind speed 
of WR 142. Assuming v$_{\perp}$ $\approx$ v$_{\infty}$ $=$ 5500 km/s 
(Kingsburgh et al. 1995), the maximum X-ray temperature from a colliding wind 
shock (eq. 1 of Luo et al. 1990) would be kT$_{\rm max}$ $\approx$ 1.95 keV 
$\mu$ [v$_{\perp}$/ (1000 km/s$^{-1}$)]$^2$ $\geq$ 59 keV for WR 142. The 
symbol $\mu$ is the mean atomic (ionic) weight per particle (ions $+$ electrons). 
This approximation does not assume specific 
WO-type wind abundances (though $\mu$ $\geq$ 1 for WR stars), but shows that 
a colliding wind shock onto an unseen companion or its wind 
could explain the high temperatures 
indicated by APEC spectral fits. 

If we assume a B0V companion with a radius R$_{*}$ $=$ 7.4 R$\sun$ (Allen 
1976), we find a separation of $\approx$ 1 AU is required to produce the 
observed {\em Chandra} X-ray luminosity through colliding wind shocks (eq. 81 
of Usov 1992). If we account for clumped winds, the separation scales as 
\.{M}$^{1/2}$ and would reduce to 0.7 AU if the mass loss rate is reduced by a factor 
of 2. At a distance of 1.23 kpc, a 1 AU separation between WR 142 and its 
hypothetical companion would produce an angular separation of much less than 
1$''$ ($\approx$ 1 mas) and would not be resolvable by {\em Chandra}. 
However, this separation is smaller than the escape radius for X-rays, which is 
375 - 1500  R$\sun$ (Section 5.1). A favorable orbital phase and geometry with 
the wind blown cavity around the companion along our line of sight would allow 
for the X-rays to escape (e.g. as occurs in $\gamma^2$ Vel; Schild et al. 2004), 
especially with clumped winds. To consider shock emission from the fast winds 
impacting circumstellar material, more information on the CS material's density, 
geometry, and distance from WR 142 would be necessary. 

\subsubsection{Magnetically Confined Wind Shocks}

For luminous stars, a surface magnetic field may be able to channel wind flow 
into shock collisions with sufficient velocity to produce hard X-ray emission. 
Generally, the magnetic field lines are thought to bend radiatively driven winds 
towards the magnetic equator, where the hemispheric winds collide. Such 
magnetically-confined wind theories have been used to explain X-ray emission 
for magnetic Ap-Bp stars (Babel \& Montmerle 1997a) and 
young O-type stars (Babel \& Montemerle 1997b, $\theta$ Ori C in Gagn\'{e} et al. 
2005). The winds were not radiatively driven in the Ap-Bp model but the resulting 
shocks are unaffected. We can determine the minimum magnetic field strength 
necessary to confine the winds in comparison to the wind magnetic confinement 
parameter using equation 7 of ud-Doula \& Owocki (2002), which is dependent 
upon the mass loss rate, v$_{\infty}$, and radius of the star. We approximate the 
radius of WR 142 to be one solar radius (Abbott et al. 1986). We find that B 
$\approx$ $\sqrt{\eta_{*}}$ 20 kG at the stellar surface. Here, $\eta_{*}$ is the wind 
magnetic confinement parameter and is 1 for marginal confinement, 10 for strong 
confinement that produces shocks strong enough to produce $\>$ 1 keV 
emission (ud-Doula \& Owocki 2002). Wind clumping could reduce the minimum 
magnetic field strength by a factor of $\sqrt{2}$ to 2. It remains to be shown that 
WR stars have such strong magnetic fields.

Additionally, it can be shown that the maximum temperature at 
the shock front from MCWS could reach maximum temperatures of tens of keV 
(eq. 4 of Babel \& Montmerle 1997a)
for high-velocity ionized metal-rich winds ($\mu$ $\geq$ 1.33)
such as that of WR 142. 
The highest temperature is only possible at the shock front and will drop off to 
lower temperatures for most of the shocked plasma, but still easily could explain 
the observed hard emission from WR 142. It is important to note though that 
MCWS theories have not yet been extended to WR stars, where several 
problems such as larger mass loss rates have not yet been addressed. 
Additionally, a magnetic rotator model with modulated the winds was invoked to 
explain the periodic variations of the X-ray emission from $\theta^{1}$ Orionis C 
(Babel \& Montmerle 1997b) but WR 142 here shows no such periodic X-ray 
emission or B-field detection so far.

\subsubsection{Coronal Emission}

As suggested for some OB stars, X-rays 
could be formed in a thin corona at the base of the wind (e.g. Cassinelli \& Olson 
1979). 
However, our calculations suggest 
that the kT $\approx$ 2 keV X-rays emerge 
from very large radii, roughly r $\gtsimeq$ 1000 R$_{*}$,
assuming a spherical homogenous wind. 
This is too far out to be attributed 
to a corona. Also, X-ray variability would be expected
with coronal emission as magnetic (coronal) emission 
is usually associated with short-term X-ray 
variability (e.g. flares). No significant variability or flaring was detected in the {\em 
Chandra} observation.

\subsection{Nonthermal Emission Processes}

Electrons in the stellar wind accelerated to relativistic energies by the Fermi 
mechanism and trapped in weak magnetic fields
could scatter stellar UV photons to produce X-ray emission  and 
$\gamma$-ray emission (Chen \& White 1991a, b), as suggested for WR stars by 
Pollock (1987). The accelerated electrons undergo adiabatic cooling between 
shocks and thus can only lose heat through nonthermal processes, such as 
inverse Compton, nonthermal bremsstrahlung, or synchrotron radio emission 
(Chen \& White 1991a). Such processes may be at work in WR 142: X-ray 
spectra of WR 142 can be fitted with a power-law and Berkeley 87 lies close to an 
extended region of high-energy $\gamma$-ray emission in Cygnus (Abdo et al. 
2007, 2009; Smith et al. 2005), though the primary source of the $\gamma$-ray 
emission has not yet been proven to be Berkeley 87 itself. 

For OB stars, the radiation will produce a power-law 
electron spectrum where the X-ray flux from inverse Compton emission should 
scale with energy as  F${_{\rm X}}$ $\propto$ E$^{-1/2}$ for electron index n $=$ 
2 (Chen \& White 1991a). Spectral fits in Table 3, when the given photon index 
$\Gamma_{\rm pow}$ $=$ 1 is converted to an energy index, give F${_{\rm X}}$ 
$\propto$ E$^{0}$, though the photon index was poorly constrained. If the photon 
index is frozen to $\Gamma$ $=$ 1.5 so as to produce Chen \& White's energy 
power-law index, the fit is reasonable but exhibits slightly more absorption and 
higher X-ray flux and luminosity (Table 3 footnotes).

The X-ray  emission  is predicted to be hard without large scale variations,  and if 
present, variations only on timescales of hours to days (Chen \& White 1991a). 
However, the Chen \& White (1991a) model is tuned to OB stars, such that 
inverse Compton is the assumed emission mechanism whereas bremsstrahlung 
may be just as important or even dominate inverse Compton for WR stars. 
Additionally, other factors, such as the coulomb effects in the dense WR wind, 
have not been fully evaluated (Chen \& White 1991a). Still, the nonthermal 
emission process is plausible for some WR stars, especially where nonthermal 
radio emission is detected (Pollock 1987). WR 142, is as of yet, undetected in the 
radio. Cappa et al. (2004) determined an upper limit of $<$0.90 mJy for the 3.6 
cm flux density from WR 142. 
Despite a lack of detection of nonthermal radio emission at 3.6 cm, radio 
observations at longer wavelengths where nonthermal emission could lead to 
higher flux densities would be useful. The hard and essentially featureless X-ray 
spectrum of WR 142 could be nonthermal, but a higher signal-to-noise X-ray 
spectrum will be needed to distinguish between thermal and nonthermal 
emission.

\section{Comments on Wolf-Rayet Star X-ray Emission}

The X-ray detection of WR 142 presents a major question: why don't we observe 
X-rays from the WC subtype when we detect hard emission from a WO star? 
When plotting L$_{\rm bol}$ versus L$_{\rm X}$, WR 142 lies in the region 
occupied by undetected WC stars (Figure 5). WOs are more evolved than WCs, 
and WO winds are even more metal-rich and should  thus be more efficient 
absorbers of soft X-rays. 

However, WR 142 has a very high terminal wind speed and a very high effective 
temperature (T$_{\rm eff}$) compared to WC stars (Crowther 2007). A higher 
effective temperature would lead to a more fully ionized wind, which lowers the 
wind opacity ($\kappa$) to X-rays (see eq. 6 of Ignace et al. 2000). The higher 
wind speed leads to a smaller radius of optical depth unity for X-rays of a given 
energy where R($\tau$ $=$ 1, E) $=$ \.{M}/(4$\pi$$v_\infty$) $\times$ $\kappa$
(E) (eq. 9 of Ignace et al. 2000). In the case for WR 142, the terminal wind speed 
is greater and the wind opacity may be smaller than for WC stars, leading to a 
smaller R($\tau$ $=$ 1, E). Thus, X-rays may be able to escape from smaller 
radii closer to the star, which could explain why WR 142 has been detected in 
the X-rays but why WC stars have (so far) not been detected. 

Another possible explanation is that WR 142 is not a single star (see Section 
5.2). Some WC stars that are members of binary systems have been known to 
display hard X-ray emission (e.g. $\gamma^2$ Vel, WC8$+$O7.5)
but the X-ray emission from WR 142 is 
much less  luminous. WR$+$O binaries display much stronger emission. 
Perhaps X-ray emission from a binary decreases as a WR star evolves from WN 
to WC to WO because of changes within the star or its wind. Maybe the 
luminosity difference lies in evolution of the unseen companion star. WR 142, if a 
binary, may not have an O star companion but some other less massive non-
degenerate stellar companion. As discussed in Section 5.4.2, a close companion 
at the right separation could account for a lower X-ray luminosity using a 
colliding wind model.

The hard X-ray emission and fast wind from WR 142 may indicate a colliding 
wind shock that could be explained by an as of yet undetected companion at 
close separation, such as a B0V star companion at $\sim$ 1 AU from WR 142 
(Section 5.4.2). This close separation would not be resolved by {\em Chandra}. If 
a less massive unseen companion is present, then {\em Chandra} may be 
observing colliding wind shock emission at or near the surface of the companion 
(Usov 1992) and no intrinsic emission from the stars themselves. Even if the undetected companion star 
had no wind itself, the overwhelming wind from WR 142 would shock onto the 
companion's surface and produce the same effect. For example, a colliding wind 
shock onto a less massive companion could occur if the companion were an 
X-ray faint B- or A-star and any intrinsic X-ray emission from WR 142, if any exists, 
were completely absorbed, e.g. by its wind.  If this is the case, then we do not 
detect X-ray emission from WR 142 itself and there is no contradiction with 
non-detections of single WC stars.  

Another thought to consider could involve the very high wind speeds of WR 142. 
What if the winds give rise to some exotic nonthermal emission processes, such 
as proposed by Chen \& White (1991a), that do not get triggered in WR stars with 
lower terminal wind speeds or lower T$_{\rm eff}$? The detection of X-rays from 
the WO type WR 142 may even suggest that the absence of X-rays from WC type 
stars be considered tentative until sensitive observations of a larger sample of 
WC stars are obtained.

\section{Conclusions}

\begin{enumerate}

\item {\em Chandra} has detected hard, heavily-absorbed X-ray emission from 
the rare WO-type star WR 142. No soft emission below 2 keV was detected by 
{\em Chandra}.

\item The observed X-ray flux is consistent with the flux from a previous {\em 
XMM-Newton} observation, within the uncertainties, and no significant X-ray 
variability was detected during the {\em Chandra} observation. 

\item Due to the faint emission from WR 142, lack of prominent emission lines, 
and low numbers of counts, statistics are unable to distinguish between thermal 
and nonthermal X-ray emission mechanisms when fitting the spectrum and both 
have been considered. If the emission is thermal, very high plasma temperatures 
are inferred.

\item In addition to WR 142, {\em Chandra} detected seven luminous OB stars in 
Berkeley 87, while eight other B stars were undetected. The hard X-rays and 
excess absorption of WR 142 contrast with two X-ray bright OB stars in Berkeley 
87, which display predominately soft X-ray emission and absorptions consistent 
with optical values. 

\item Though the X-ray emission mechanism in WR 142 is unclear, the hard 
X-ray spectrum observed by {\em Chandra} could be explained by a colliding wind 
shock onto an as yet unseen companion, though the escape of the X-rays may 
require a geometry with the companion in front of WR 142. A colliding wind 
shock interpretation would also resolve an apparent contradiction of non-detections 
in the X-rays of single WC stars while WR 142 (a WO star) was 
detected. A key point not to be overlooked is that
colliding winds can produce very hot plasma (as observed)
even in the absence of magnetic fields.

\item Alternatively, mechanisms that assume stellar 
magnetic fields such as MCWS or inverse Compton 
scattering (as formulated by Chen \& White 1991a) 
could play a role in the X-ray emission 
of WR 142. However for MCWS, very strong 
surface fields of tens of kG would be required to 
effectively confine the powerful WR 142 wind (Sec. 5.4.3), 
also noted  by Oskinova et al. (2009) 
in their MCWS interpretation. There is so far no 
observational evidence of such strong B-fields in 
WR stars, nor do the existing X-ray data show 
any obvious signatures of impulsive X-ray flares 
that often accompany magnetic reconnection. 
If WR 142 has even a weak surface magnetic field, then 
X-ray production via inverse Compton scattering 
provides an attractive alternative to the extreme 
conditions required for magnetic wind confinement. 
A key question is whether the X-ray spectrum of 
WR 142 is indeed a power-law, as expected for 
inverse Compton scattering. To answer this question, 
a higher signal-noise X-ray spectrum from a much 
deeper observation will be required.

\end{enumerate}

\acknowledgments

This work was supported by {\em Chandra} award GO9-0004X issued by the 
Chandra X-ray Observatory Center (CXC). The CXC is operated by the 
Smithsonian Astrophysical Observatory (SAO) for, and on behalf of, the National 
Aeronautics Space Administration under contract NAS8-03060.
We have utilized NASA's Astrophysics Data
System Service and the SIMBAD database. This research has made use of data 
obtained from the High Energy Astrophysics Science Archive Research Center 
(HEASARC), provided by NASA's Goddard Space Flight Center.

\clearpage
\begin{deluxetable}{lll}
\tabletypesize{\scriptsize}
\tablewidth{0pc}
\tablecaption{Stellar Properties of WR 142}
\tablehead{
\colhead{Parameter}      &
\colhead{Value}  &
\colhead{Notes\tablenotemark{a}}     
}
\startdata
Spectral Type                             		& WO2      				& (1)(2) \nl
Age (Myr)						&  $\sim$3 - 5 			& (3) \nl
d (pc)                              	 		& 1230$\pm$40                  & (4) \nl 
V (mag.)                             			& 13.4                              	& (2)  \nl
A$_{\rm V}$ (mag)                        	& 6.1                        		& (5) \nl
J,H,K$_{\rm s}$ (mag)                        	& 9.54,8.89,8.60                 & (6) \nl
log M/M$_{\sun}$            			& 0.9$\pm$0.2           		& (7) \nl
\.{M} (M$_{\sun}$ yr$^{-1}$)               	& 1.7 $\times$ $10^{-5}$	& (8) \nl
v$_{\infty}$ (km s$^{-1}$)  		& 5500$\pm$200  		& (2) \nl
log L$_{\rm wind}$ (ergs s$^{-1}$)	& 38.2          			&  (9) \nl
log L$_{\rm bol}$/L$_{\sun}$          	& 5.35, 5.65         		& (10) \nl

\enddata
\tablenotetext{a} {Refs. and Notes:
(1) Crowther et al. (1998)~
(2) Kingsburgh et al. (1995)~
(3) Massey et al. (2001), Polcaro \& Norci (1998), Turner et al. (2010)~
(4) Turner et al. (2006)~
(5) van der Hucht 2001, using A$_{\rm V}=$0.9A$_{v}$~
(6) 2MASS Point Source Catalog~
(7) Smith et al. (1994)~
(8) Unclumped value from Barlow (1991), Clumping could reduce \.{M} by a factor of $\approx$ 2 - 4 (Crowther 2007) ~
(9) Polcaro et al. (1991), also L$_{\rm wind}=$(1/2)\.{M}v$_{\infty}^2$~
(10) Determined by fitting photometric observations using a range of clumped mass loss rates, Oskinova et al. 2009~
}
\end{deluxetable}
\clearpage

\begin{deluxetable}{lllllllll}
\tabletypesize{\tiny}
\tablewidth{0pt} 
\tablecaption{X-ray Properties of WR 142 and OB stars\tablenotemark{a}  in Berkeley 87}
\tablehead{
	 \colhead{Name}	&
           \colhead{R.A.}               &
           \colhead{Decl.}                             &
           \colhead{Net Counts}                   &
           \colhead{E$_{25}$,E$_{50}$,E$_{75}$} &
           \colhead{$<$E$>$}                                       &
           \colhead{P$_{\rm const}$}                                    &
           \colhead{Sp. Type}  &
           \colhead{Identification (offset)}                                    \\    
           \colhead{}	&
           \colhead{(J2000)}                         &
           \colhead{(J2000)} &
           \colhead{(cts)}                                          &               
           \colhead{(keV)}                                          &                  
           \colhead{(keV)}                                          &         
           \colhead{KS/GL}                                          & 
           \colhead{}  &
           \colhead{(arcsec)} 
                                  }
\startdata

HD 229059		& 20 21 15.37 & $+$37 24 31.2 & 341$\pm$19 	& 1.09,1.36,1.67 	& 1.54 	& 0.40/0.93 	& B1 Ia\tablenotemark{b}  	& HST 202115.36$+$372431.1 (0.16)\tablenotemark{c} \\
Berkeley 87-4		& 20 21 19.26 & $+$37 23 24.6 & 63$\pm$8 		& 0.96,1.19,1.67 	& 1.53 	& 0.10/0.49 	& B0.2 III 					& HST 202119.25$+$372324.7 (0.16)\\
Berkeley 87-13		& 20 21 31.60 & $+$37 20 40.6 & 5$\pm$2 		& ...,1.25,... 		& 1.60 	& 0.55/... 		& B0.5 III 					& HST 202131.55$+$372041.4 (1.00)\\
Berkeley 87-24		& 20 21 38.21 & $+$37 25 17.4 & 9$\pm$3 		& 1.59,1.99,3.49 	& 2.41 	& 0.49/... 		& B1 Ib 					& HST 202138.33$+$372516.3 (1.80)\\
BD$+$36 4032		& 20 21 38.69 & $+$37 25 15.4 & 254$\pm$16 	&1.08,1.34,1.64 	& 1.43 	& 0.01/0.62 	& O8.5 III\tablenotemark{b}  	& HST 202138.68$+$372515.2 (0.23)\\
Berkeley 87-26		& 20 21 39.72 & $+$37 25 05.6 & 11$\pm$3 		& 0.92,1.09,1.78 	& 2.00 	& 0.70/... 		& B0.5 Iab 				& HST 202139.75$+$372505.9 (0.47)\\
WR 142 			& 20 21 44.35 & $+$37 22 30.7 & 46$\pm$7		& 2.99,4.28,5.32 	& 4.31 	& 0.77/0.40 	& WO2\tablenotemark{d} 		& HST 202144.36$+$372230.5 (0.23)\\
Berkeley 87-32		& 20 21 47.35 & $+$37 26 32.2 & 11$\pm$3 		& 0.95,1.09,1.57 	& 1.33 	& 0.91/... 		& B0.5 III 					& HST 202147.35$+$372632.0 (0.18)\\

\enddata
\tablecomments{
X-ray data using events in the 0.3 - 8 keV range within a 3$\sigma$ {\em wavdetect} ellipse, which for WR 142 is $\approx$ 11.25 pix$^{2}$. 
Tabulated quantities are: J2000.0 X-ray position (R.A., decl.); net counts and 
net counts error from {\em wavdetect} (accumulated in a 70148 s exposure, rounded to the nearest integer,
background subtracted and PSF-corrected); 25$\%$, 50$\%$, and 75$\%$ photon quartile energies (E$_{25}$, E$_{50}$, and E$_{75}$), 
mean photon energy ($<$E$>$); probability of constant count-rate determined by the Kolmogorov-Smirnov (KS) 
test and the Gregory-Loredo (GL) algorithm, where P$_{\rm const}$ $\leq$ 0.05 would indicate likely variability;
spectral type as indicated by SIMBAD or otherwise noted; and HST GSC candidate counterpart identification
within a 2$''$ search radius. The offset (in parentheses) is given in arc seconds between the X-ray and 
counterpart position. Ellipses indicate insufficient counts for reliable measurement. }

\tablenotetext{a}{Faint B stars not detected:
   Berkeley 87 nos. 7 (B), 9 (B0.5 V), 15 (V439 Cyg, B0e/B[e] - references in Sec. 4.3), 16 (B2 V), 18 (B1 V), 
                    31 (B1 V), 34 (B), and 38 (B2 III). Spectral types are from the SIMBAD database.}
                    
\tablenotetext{b}{Spectral type from Massey et al. (2001).}

\tablenotetext{c}{Two close HST sources have nearly the same sexagesimal coordinates and are indistinguishable by {\em Chandra}.}

\tablenotetext{d}{Spectral type references in Table 1.}

\end{deluxetable}

\clearpage

\begin{deluxetable}{llllll}
\tabletypesize{\scriptsize}
\tablewidth{0pc}
\tablecaption{{\em Chandra} Spectral Fits for WR 142
   \label{tbl-1}}
\tablehead{
\colhead{Parameter}      &
\colhead{       }        &
\colhead{       }        &
\colhead{       } 
}
\startdata
Model                             							& 1T APEC   					&   POWER			& POWER $+$ GAUSS  \nl
N$_{\rm H}$ (10$^{22}$ cm$^{-2}$) 					& 5.66						& 4.61				& 4.55		\nl
kT (keV)                   		 						& \{20.\}\tablenotemark{a} 		& ...					& ...			\nl
norm (10$^{-13}$)          	  						& 6.36						& ...					& ...			\nl
$\Gamma_{\rm pow}$\tablenotemark{b}                          	& ...							& 1.0	\tablenotemark{c}	& \{1.0\}		\nl
norm$_{\rm pow}$ (10$^{-6}$) 						& ...							& 2.31				& 2.32		\nl
E$_{\rm line}$ (keV)                  						& ... 							& ...					& \{6.67\}		\nl
$\sigma_{\rm line}$ (keV)             					& ...							& ...					& \{0.05\}		\nl
norm$_{\rm line}$ (10$^{-8}$)       					& ...							& ...					& 8.68		\nl                        
Abundances\tablenotemark{d}                    			& WO						& ...					& ...			\nl
C-statistic\tablenotemark{e}						& 215.90						& 215.08				& 214.73		\nl
F$_{\rm X}$ (10$^{-14}$ ergs cm$^{-2}$ s$^{-1}$)		& 1.65 (3.66)					& 1.74 (2.93)			& 1.77 (2.96)	\nl
F$_{\rm X,line}$ (10$^{-16}$ ergs cm$^{-2}$ s$^{-1}$)	& ...							& ...					& 8.66 (9.28)	\nl
log L$_{\rm X}$ (ergs s$^{-1}$)   		 			& 30.82						& 30.72				& 30.	73		\nl
\enddata
\tablecomments{
Based on  fits of ACIS spectra using XSPEC v12.5.0. 
The data were unbinned and not background subtracted (background is negligible). 
The tabulated parameters
are absorption column density (N$_{\rm H}$), plasma temperature (kT),
XSPEC normalization (norm), photon power-law index ($\Gamma_{\rm pow}$), power-law model
normalization (norm$_{\rm pow}$),
Gaussian line centroid energy (E$_{\rm line}$), line width
($\sigma_{\rm line}$ $=$ FWHM/2.35), line normalization (norm$_{\rm line}$).
Quantities enclosed in curly braces were held fixed during fitting.
WO abundances are from Table 1 of van der Hucht et al. (1986), 
except the hydrogen and nitrogen abundances are arbitrarily set to the small non-zero value 
1 $\times$ 10$^{-6}$ for compatibility with XSPEC. 
Solar values from Anders \& Grevesse (1989) are used for any elements in the XSPEC abundance 
table that are not listed in van der Hucht et al. (1986). 
X-ray flux (F$_{\rm X}$) is the  observed (absorbed) value followed
in parentheses by the unabsorbed value in the 0.3 - 8.0 keV range.
The continuum-subtracted Gaussian line flux (F$_{\rm X,line}$) 
is measured in the 6.5 - 6.9 keV range. 
X-ray luminosity (L$_{\rm X}$) is the unabsorbed value in the
0.3 - 8.0 keV range. A distance of 1230 pc is assumed (Turner et al. 2006).} 
\tablenotetext{a}{If kT is allowed to vary, it runs away to the maximum temperature of 64. keV allowed by the 
APEC model. If frozen at a higher temperature of kT $=$ 50 keV, 
the fit is similar and converges to N$_{\rm H}$ $=$ 5.38 $\times$ 10$^{22}$ cm$^{-2}$, C-statistic $=$ 
215.56, F$_{\rm X}$ $=$ 1.67 (3.43) $\times$ 10$^{-14}$ ergs cm$^{-2}$ s$^{-1}$, and log L$_{\rm X}$ $=$ 30.79 ergs s$^{-1}$.}
\tablenotetext{b}{A photon index $\Gamma_{\rm pow}$ equates to F$_{\rm X}$ $\propto$ E$^{-(\Gamma_{\rm pow} -1)}$.
      That is, values of $\Gamma_{\rm pow}$ $>$ 1 correspond to flux decreasing with
      increasing energy E.}
\tablenotetext{c}{If the photon index is frozen to $\Gamma_{\rm pow}$ $=$ 1.5, the fit gives N$_{\rm H}$ 
$=$ 5.91 $\times$ 10$^{22}$ cm$^{-2}$, C-statistic $=$ 215.73, F$_{\rm X}$ $=$ 1.64 (4.03) $\times$ 10$^{-14}$ 
ergs cm$^{-2}$ s$^{-1}$, and log L$_{\rm X}$ $=$ 30.86 ergs s$^{-1}$.}
\tablenotetext{d}{Varying the abundances did little to improve fits.}
\tablenotetext{e}{Regrouped spectra to a minimum of 6 counts per bin enabled the use of $\chi^2$ statistics. 
For 1T APEC models, this yielded a lower 90\% confidence
bound on kT$>$3.6 keV. 90\% bounds on N$_{\rm H}$ were in the range [2.8 - 9.6] $\times$ 10$^{22}$ 
cm$^{-2}$ for models frozen at kT $=$ 20 keV. We were not able to find stable confidence 
bounds on $\Gamma_{\rm pow}$ for the power-law models.}

\end{deluxetable}

\clearpage

\begin{deluxetable}{llll}
\tabletypesize{\scriptsize}
\tablewidth{0pc}
\tablecaption{{\em Chandra} Spectral Fits for X-ray Bright OB Stars
   \label{tbl-1}}
\tablehead{
\colhead{Parameter}      &
\colhead{BD$+$36 4032}   &
\colhead{HD 229059}      &
\colhead{Berkeley 87-4}     
}
\startdata
Model                             						& 1T APEC   		& 1T APEC            	& 1T APEC   		\nl
Abundances                                      				& solar                 	& solar              		& solar     			\nl
N$_{\rm H}$ (10$^{22}$ cm$^{-2}$) 				& 1.13 [0.99 - 1.45]	& 1.16 [1.01 - 1.24]	& 0.82 [0.59 - 1.16]	\nl
kT (keV)                   		 					& 0.56 [0.35 - 0.69] 	& 0.64 [0.47 - 0.81]	& 0.54 [0.27 - 0.74]	\nl
norm (10$^{-4}$)          	  					& 1.16 [0.69 - 4.31]	& 1.47 [0.76 - 1.75]	& 0.20 [...  - 1.51]	\nl
$\chi^2$/dof                                    				& 23.7/18               	& 30.7/28            	& 2.8/4              		\nl
$\chi^2_{\rm red}$                                  			& 1.32                  	& 1.10           		& 0.70              		\nl
F$_{\rm X}$ (10$^{-13}$ ergs cm$^{-2}$ s$^{-1}$) 	& 0.20 (2.91)		& 0.32 (3.93)  	     	& 0.05 (0.53)       	\nl
log L$_{\rm X}$ (ergs s$^{-1}$)   	         			& 31.72			& 31.85	             	& 30.98              		\nl
\enddata

\tablecomments{
Based on  fits of ACIS spectra binned to a minimum of 10 counts per bin using XSPEC v12.5.0.
Square brackets enclose 90\% confidence intervals. An ellipsis means the algorithm 
used to compute 90\% confidence limits did not converge.
The X-ray flux is the absorbed value, followed in parentheses by the unabsorbed value.
A distance of 1.23 kpc is assumed. Solar abundances are from Anders \& Grevesse (1989).
}

\end{deluxetable}

\clearpage

\begin{figure}
\figurenum{1}
%\begin{tabular}{cc}
\includegraphics*[width=5.8cm,angle=0]{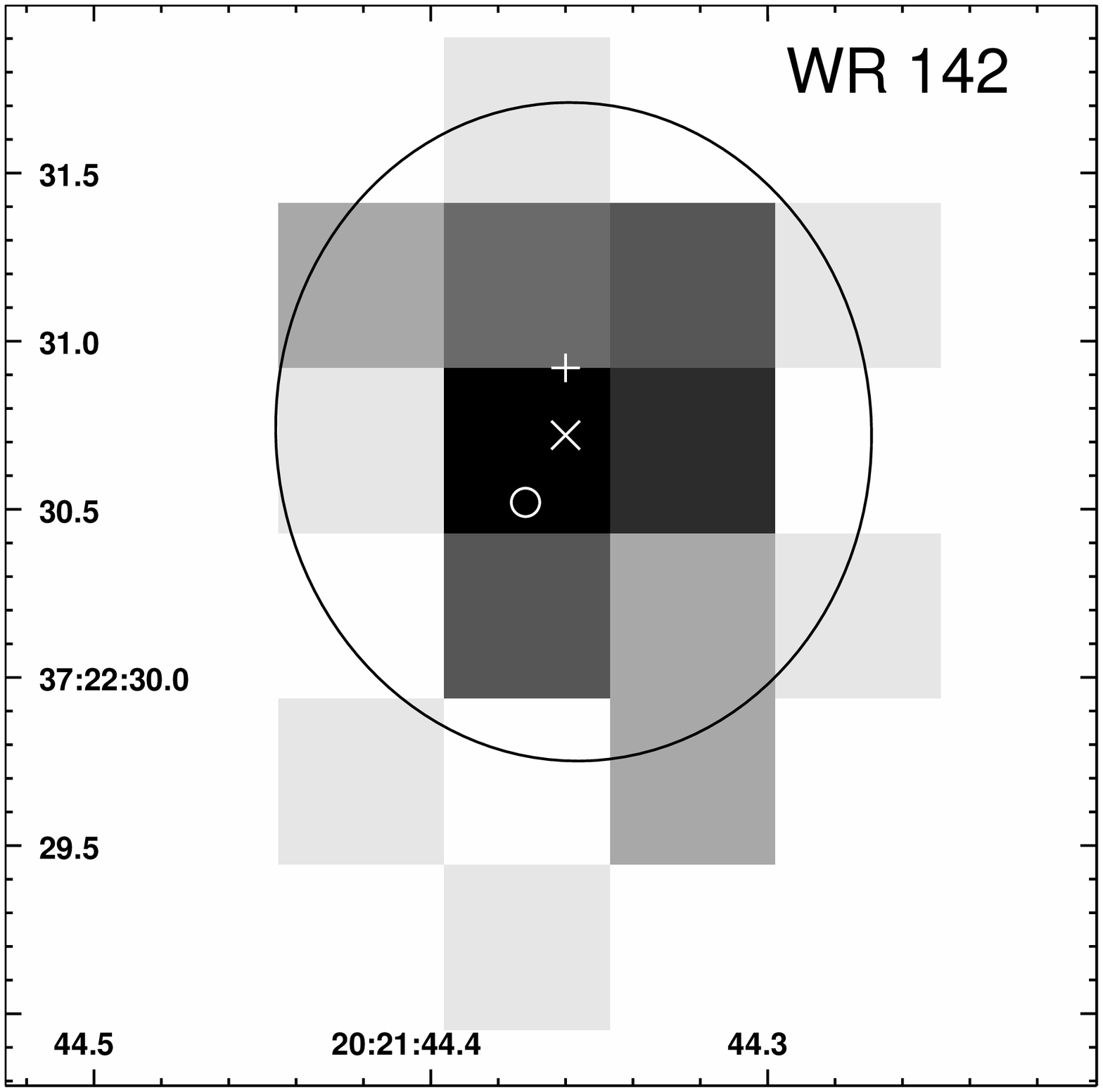}
\includegraphics*[width=12cm,angle=0]{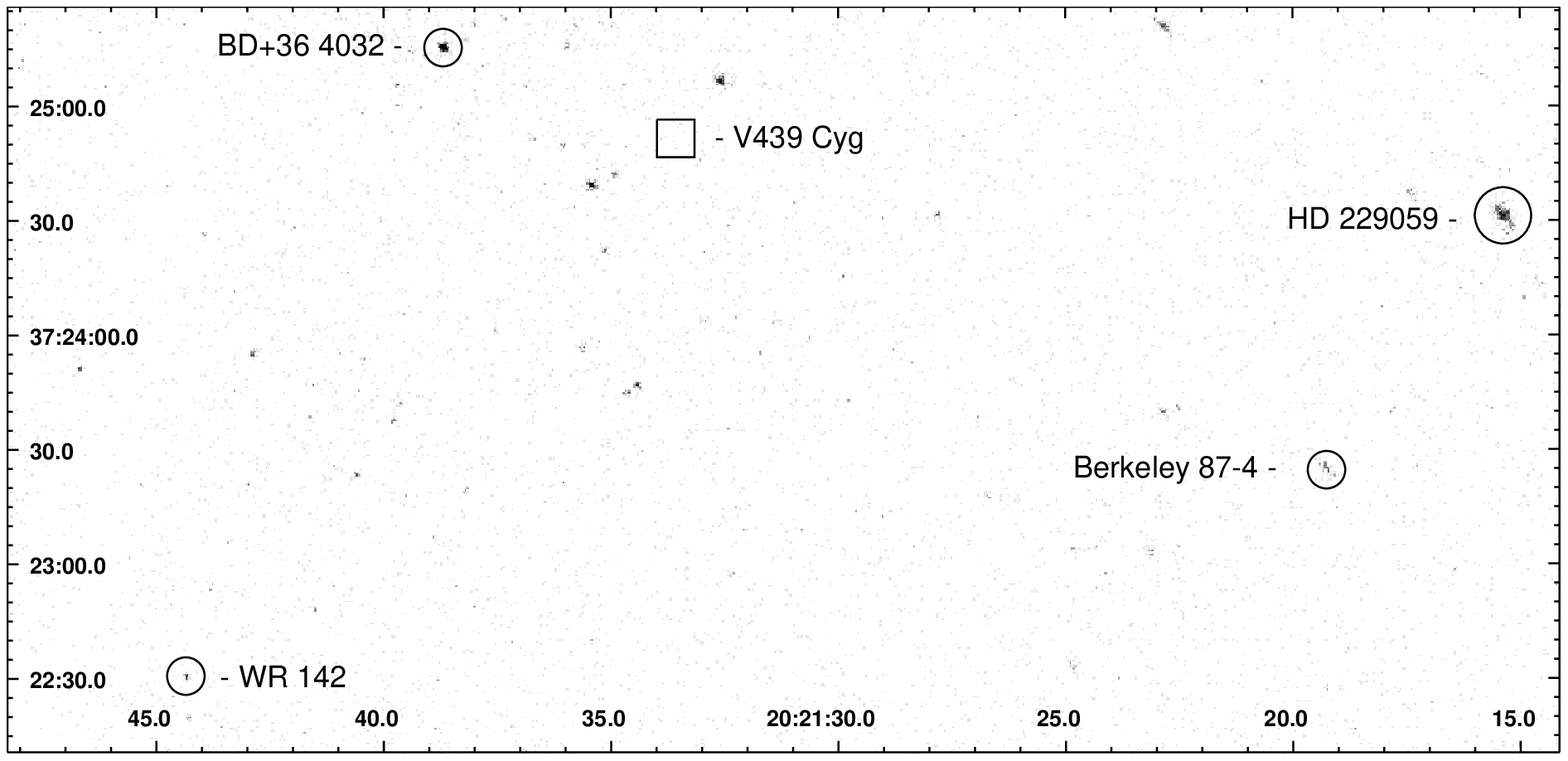}
%\end{tabular}
\caption{Left: An ACIS-I image (0.3 - 8 keV, log intensity scale) centered near WR 142. 
Indicators give the WR 142 position as follows: circle marks HST GSC 2.3.2 (optical), plus 
marks USNO-B1.0 (optical), and the cross marks both the {\em Chandra} and 2MASS 
(near-infrared) positions. WR 142 was 
only 28$''$ from the aimpoint. The 3$\sigma$ ellipse determined by {\em wavdetect} is shown 
and encloses 46 $\pm$ 7 net counts. Right: An ACIS-I wide field image (0.3 - 8 keV, log 
intensity scale) showing the X-ray bright OB stars relative to WR 142, where circles mark 
the {\em Chandra} positions and the square marks a 2MASS position for the non-detection 
of V439 Cyg. All coordinates are J2000.0 and pixels are 0.492$''$. }
\end{figure}

\clearpage

\begin{figure}
\figurenum{2}
\includegraphics*[width=9cm,angle=-90]{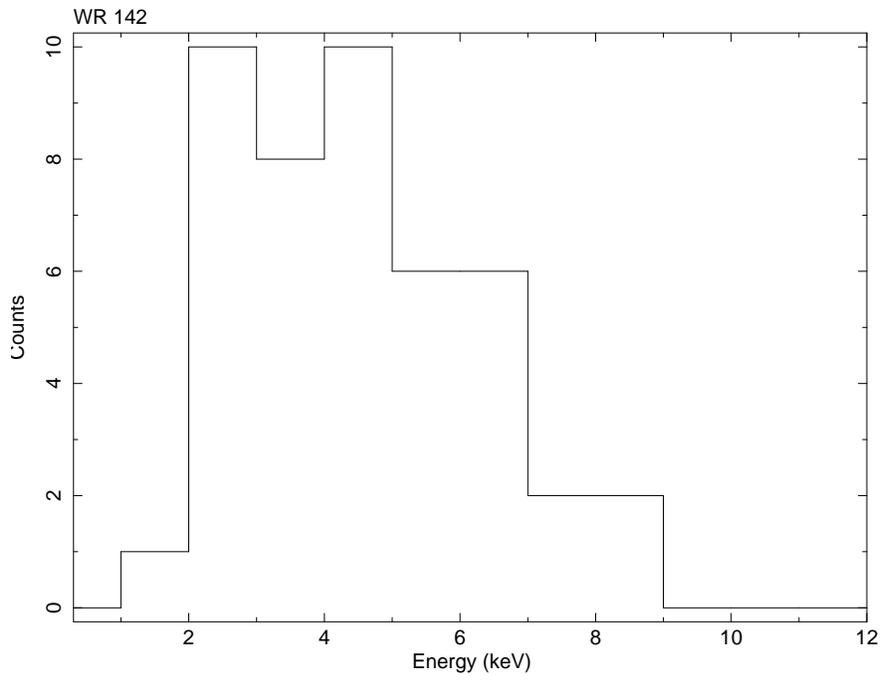}
\caption{A histogram showing the observed X-ray event energies for WR 142 in 
the energy range from 0.3 - 12 keV ({\em wavdetect} 3$\sigma$ ellipse extraction region). The bin size is 1 keV.}
\end{figure}

\clearpage

\begin{figure}
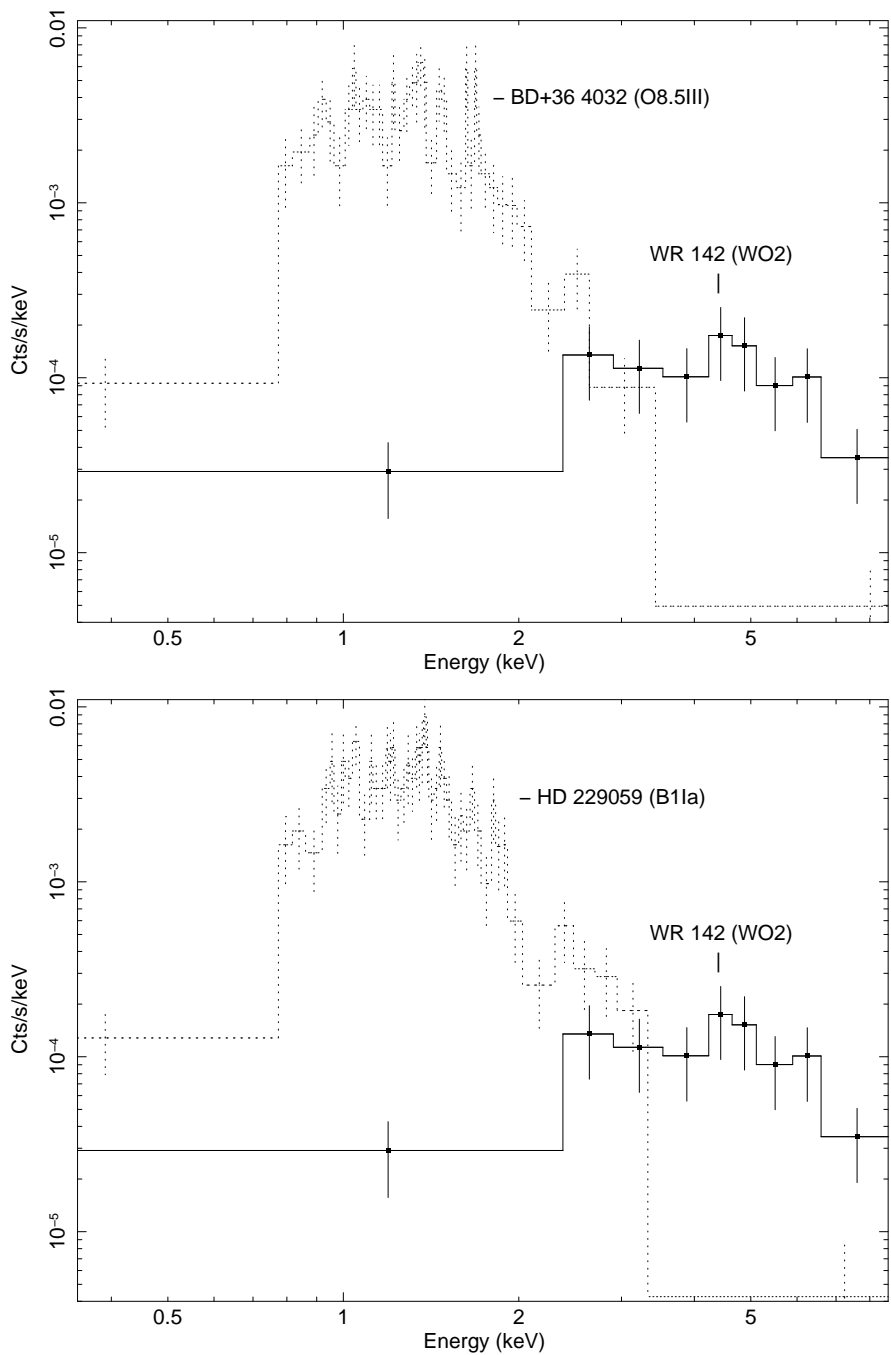

\figurenum{3}
\includegraphics*[width=9cm,angle=-90]{f3_top.ps}
\includegraphics*[width=9cm,angle=-90]{f3_bottom.ps}
\caption{{\em Chandra} ACIS-I spectra of WR 142 (46 counts) and OB stars, top: the O8.5 III
          star BD$+$36 4032 (249 counts), bottom: the B1 Ia star
          HD 229059 (341 counts). Spectra were rebinned to a minimum of 5 counts
          per bin. The straight lines between data points
          are a histogram and not a fitted model. The
          spectra show a clear contrast. The WR 142 spectrum
          is dominated by hard emission above 2 keV
          while soft emission below 2 keV dominates
          the OB stars.}
\end{figure}

\clearpage

\begin{figure}
\figurenum{4}
\includegraphics*[width=9cm,angle=-90]{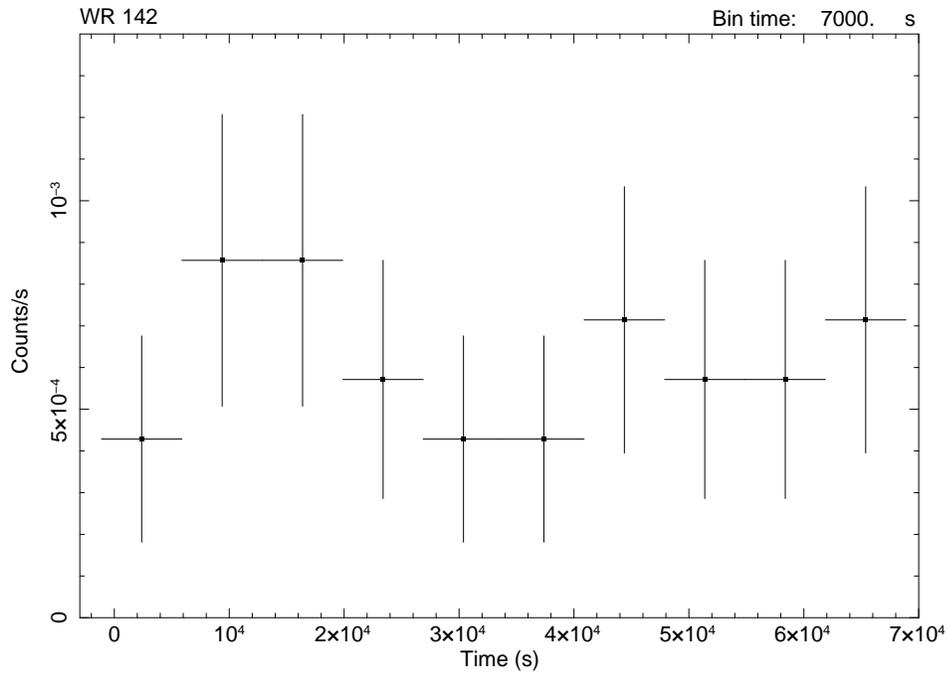}
\caption{A {\em Chandra} lightcurve for WR 142 in the 0.3 - 8 keV energy range 
(1$\sigma$ error bars). The extraction region was a 3$\sigma$ ellipse (Figure 1) 
given by {\em wavdetect} output.  }
\end{figure}

\clearpage

\begin{figure}
\figurenum{5}
\includegraphics*[width=9cm,angle=-90]{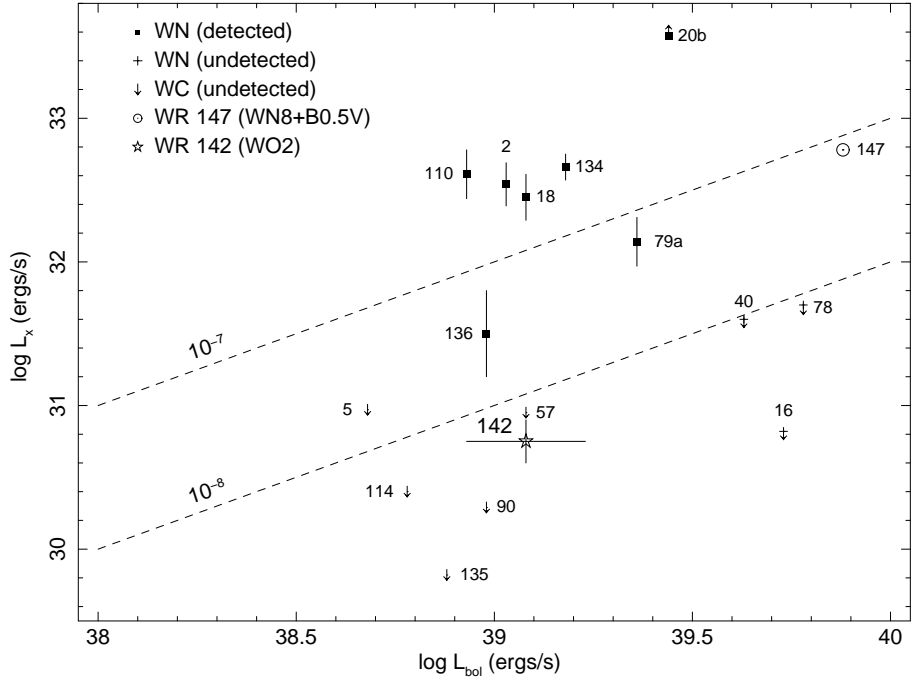}
\caption{Adapted from Skinner et al. (2010b), with new points for 
WR 142 and WR 16. Note that WR 142 lies
     in the region of undetected WC stars. The dashed line marked
     10$^{-7}$ approximately follows the L$_{\rm X}$ 
     versus L$_{\rm bol}$ relation for O- and early B-stars from {\em ROSAT} 
     (Bergh\"{o}fer et al. 1997), though a large scatter exists. The  L$_{\rm X}$
     value   for WR 16  is based on  a recent  {\em XMM-Newton}
     observation (Observation Id  0602020301, obtained on
     2009 Dec. 28). EPIC MOS images show a faint excess above
     background at the WR 16 position (13 net MOS2 counts in the
     0.3 - 8 keV range, r $=$ 15$''$  extraction circle $=$ 68\%
     encircled   energy,  33 ksec of usable MOS2 exposure,
     0.394 counts ksec$^{-1}$). Using PIMMS, the MOS2 count
     rate equates to an unabsorbed X-ray luminosity
     log L$_{\rm X}$(0.3 - 8 keV) = 30.82 ergs s$^{-1}$, assuming
     d = 2.73 kpc  and A$_{\rm V}$ = 1.85
     (van der Hucht 2001), and an intrinsic 2T Raymond-Smith
     thermal plasma source spectrum with
     kT$_{1}$ = 0.6 keV and kT$_{2}$ = 3.5 keV. Because the
     X-ray detection of WR 16 is of low significance, its
     L$_{\rm X}$ value is plotted as an upper limit.
     L$_{\rm bol}$ for WR 16 is from Hamann et al. (2006). }
\end{figure}

\end{document}